\documentclass[a4paper, amsfonts, amssymb, amsmath, reprint, showkeys, nofootinbib, twoside]{revtex4-1}
\usepackage[english]{babel}
\usepackage[utf8]{inputenc}
\usepackage[colorinlistoftodos, color=green!40, prependcaption]{todonotes}
\usepackage{graphicx}
\begin{document}

\title{The Photon Hall Pinwheel\\  Radiation of Angular Momentum by a Diffusing Magneto-optical Medium}
\author{B.A. van Tiggelen}
    \email[Correspondence email address: ]{bart.van-tiggelen@grenoble.cnrs.fr}
    \affiliation{Univ. Grenoble Alpes, CNRS, LPMMC, 38000 Grenoble, France}
\author{G.L.J.A. Rikken }
    \affiliation{Univ. Grenoble Alpes, CNRS, LNCMI, EMFL, Toulouse/Grenoble, France}
\date{\today} 

\begin{abstract}
We consider an optically thick spherical agglomerate of magneto-optical scatterers with a central isotropic, unpolarized light source,
placed in a homogeneous magnetic field.
 The Photon Hall Effect
induces a rotating Poynting vector, both inside and outside the medium.
We show that electromagnetic (orbital) angular momentum leaks out and
induces a torque proportional to the injection power of the source and the photon Hall angle.
This effect represents a novel class of optical phenomena, generating angular momentum from diffusive magneto-transport.
\end{abstract}


\maketitle

\section{Introduction}
The Photon Hall Effect (PHE) was predicted \cite{phetheo1} and  measured \cite{pheexp1} more than 20 years ago. It refers to the
electromagnetic flux that is scattered in a preferential direction perpendicular to both incident current and magnetic field, much in the spirit
of the (anomalous) Hall effect in electronic conduction. The PHE has been shown to originate from Faraday rotation in single scattering from a dielectric Mie
sphere \cite{david} and vanishes in the regime of pure electric dipole coupling (the Rayleigh regime). Therefore,  the PHE does not occur in single scattering
of light by atoms, but is in that case generated
by multiple scattering \cite{delande}, or when
the electric-dipole transition interferes with a higher multipole \cite{pheatom}.
In recent literature many more or less related effects have been identified, such as the photon spin Hall effect \cite{OSpinHall,PSE,SHE}, the
quantum spin Hall effect of light \cite{bliokh}, the phonon Hall effect \cite{phononhall}, the plasmon Hall effect \cite{plasmonhall}
and even other photon Hall effects \cite{otherPHE}.

In a scattering medium with a central light source, subject to a homogeneous magnetic field $\mathbf{B}_0$ along the $z$-axis, the PHE emerges as a  current rotating around the field lines.
The Poynting vector associated with the PHE
is given by $\mathbf{S}_{\mathrm{PHE}}= D_H
\widehat{\mathbf{B}}_0\times \nabla \rho(\mathbf{r},t)$, with $\rho(\mathbf{r},t)$ the electromagnetic energy density and $D_H(B_0)$ the Hall diffusion
constant,  whose sign is determined by sense of the Faraday rotation.
The simplest case is to consider a point source $P(\mathbf{r},t) = P(t) \delta(\mathbf{r})$ injecting the power $P$  into an infinite diffuse medium
with diffusion constant $D$.
For a single shot of energy $W$, $P(t) = W\delta(t)$, we can insert the well-known solution of the diffusion equation to obtain,
\begin{equation}\label{inf}
  \mathbf{S}_{\mathrm{PHE}}(r,t) = - \frac{W D_H r}{16\pi^{3/2} (Dt)^{5/2} } \exp \left( -\frac{r^2}{4Dt} \right) \widehat{\mathbf{\phi}}
\end{equation}
whose maximum propagates outwards as $r \sim \sqrt{Dt}$ while decaying in time as $1/(Dt)^{3/2}$. The existence of such a rotating electromagnetic
current is, on its own already, surprising. Standard textbooks on electromagnetism \cite{jackson,bw,feynman} insist on the ambiguity in
the interpretation of $\mathbf{S}= c_0 \mathbf{E} \times \mathbf{H}$ as the energy current density,
based on the conservation law $\partial_t \rho + \nabla\cdot \mathbf{S}= P$ with the obvious freedom to add the curl of an arbitrary vector field,
as is the  case for the PHE.
In homogeneous, conservative media, the directions of Poynting vector
and group velocity coincide and the ambiguity is lifted \cite{LL}.
In \emph{absorbing }media subject to a magnetic field the addition of a curl turned out to be necessary \cite{curlabs}.
In \emph{inhomogeneous}, conservative media this ambiguity has so far never been addressed.
In our simple spherical model with elastic multiple scattering, the circulating light does not, \emph{in fine},  move around energy, and arguably,
there is no ambiguity. Yet this same light carries momentum that can be
transferred to matter and is thus observable \cite{AMAM}. Indeed,
the interpretation of  electromagnetic angular momentum (for $\mu=1$),
$\mathbf{K}(t)= c_0^{-2}\int d^3\mathbf{r} \ \mathbf{r} \times \mathbf{S} $,
 is not  subject to this curl ambiguity, since adding any curl to $\mathbf{S}$ would produce an additional finite angular momentum. Radiative angular momentum
 in matter  is part of another
 controversy \cite{AMAM} which is of minor importance inside our dilute scattering medium, and of no importance outside.
 The PHE generates the angular
 momentum,
\begin{equation}\label{am1}
   \mathbf{ K}(t) = -\frac{ 2 D_H}{c_0^2}\widehat{\mathbf{z}}\int d^3 \mathbf{r } \rho(\mathbf{r},t) = -\frac{ 2 D_H}{c_0^2}\widehat{\mathbf{z}} \int_0^t dt' P(t')
\end{equation}
proportional to the total amount of energy injected, and conserved in time after the injection stops.
The total torque on the diffuse medium is $\mathbf{N} = - d\mathbf{K}/dt =
2 D_H \widehat{\mathbf{z}} P(t)/c_0^2$. This torque is of course very small. For a PHE angle of
$D_H/D \sim 10^{-3}$, the largest we have realized, a mean free path of 10 $\mu$m ( $D \approx c_0 \ell/3 = 1000$ m/s$^2$) and a power of 10 W,
we find a torque $N= 10^{-16} $ Nm. The conservation of
total angular momentum can be expressed as
\begin{eqnarray}\label{AMcon}
    \frac{d}{dt}\left(  \mathbf{K} +  \mathbf{K}_{\mathrm{mat}} \right)_i
  = \lim_{r\rightarrow \infty} \int d^2A(r) \hat{r}_n
    \epsilon_{ijk} \hat{{r}}_j T_{nk}({\mathbf{r}},t)   \nonumber \\
    \end{eqnarray}
with $ \mathbf{K}_{\mathrm{mat}}  $ the angular momentum of the matter in the sphere and $\mathbf{T}$ the symmetrical electromagnetic stress tensor \cite{jackson}.
The term on the righthand side describes the flow of angular momentum to infinity. In typical scattering problems the far field is characterized by
  $T_{ij}(\mathbf{r}) = P \delta_{ij} + Q\hat{r}_i\hat{r}_j $, so that this flow vanishes
trivially.
In the presence of the PHE however, it does not.  Nevertheless, if we choose $r > c_0t$, the leak of angular momentum  is outside the light cone and
vanishes at infinity. As a result, $\mathbf{K} + \mathbf{K}_{\mathrm{mat}}$ must be conserved in time. They can exchange mutually  via
the Abraham torque $\mathbf{N}_A=  c_0^{-1}\int d^3\mathbf{r} \  \mathbf{r} \times \partial_t (\mathbf{P}\times\mathbf{ B}) $
\cite{BartPRA}.

\begin{figure}
  \includegraphics[width=7cm]{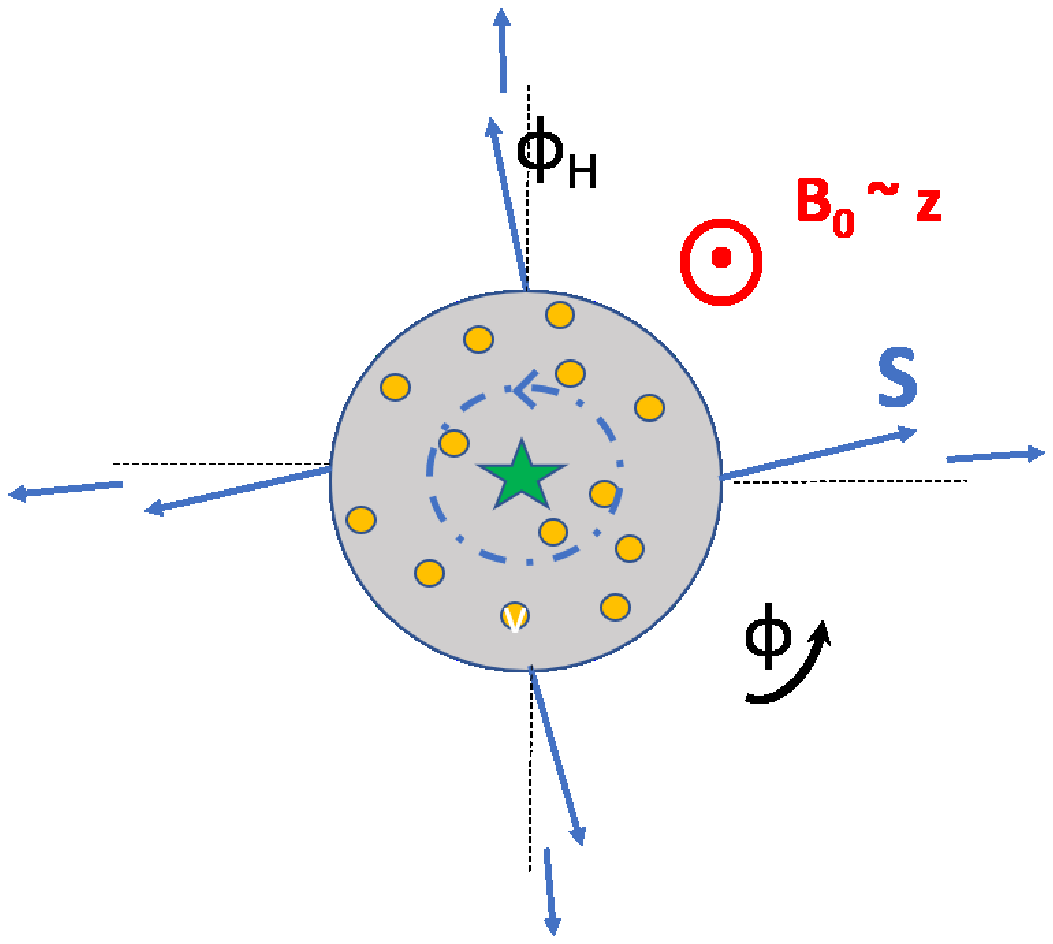}
  \caption{The Photon Hall Pinwheel: The Photon Hall Effect inside a diffuse sphere emerges as a  Poynting vector circulating around the origin,
  not affecting  energy transport
  but carrying a
  finite orbital angular momentum.
  The radiation leaves the boundaries under the Photon Hall angle $\phi_H= D_H/D$ with the local normal (here drawn for $D_H < 0$). This angle decreases as $1/r$ with distance.}\label{PHEsphere}
\end{figure}

In order to calculate the photon Hall torque on a scattering agglomerate, we  consider a finite sphere with radius $a$ (see Figure 1). Inside the sphere the diffusion equation applies as before, and for times
 $t < a^2/4D$ after the source emitted the situation is essentially the one of an infinite medium. At later times we have to know how photons leave the sample.
 In an intuitive  approximation, we can imagine the PHE at the boundaries making the photons leave the sphere
 at the "Photon Hall Angle" $\phi_H = D_H/D$ with the local normal. If
 $ P(a, t)= 4\pi a^2 S(a,t) $ is the total current leaving the sphere, the rate of angular momentum that flows outside is $a \times \phi_H P(a,t)/c_0$.
 The torque can thus be estimated as
 $N (t) \approx \phi_H P(a,t) a /c_0$. Energy conservation imposes that $P(a,t) = P(t) - \partial_t W$, with $W$ the electromagnetic energy inside the sphere.
 If we assume
 the latter is stationary, we obtain,
 \begin{equation}\label{torque1}
    N(t) = \frac{D_H}{D} \frac{a}{c_0} P(t)
 \end{equation}

 To get   quantitative insight  how angular momentum is radiated  by the sphere  into space we will follow the standard method to deal with
 boundaries of the diffusion
 equation, and calculate the electromagnetic angular momentum outside the sphere, rather than the flow of angular momentum.
 A good treatment of the skin layer  \cite{skin} is essential to  respect flux conservation. The magneto-optics of the effective medium is described by a
 complex index of refraction $\varepsilon_\sigma(\mathbf{k})  $  in terms of which wave vector $k_\sigma$ and extinction length $\ell_\sigma$ are defined as
 $\varepsilon_\sigma(\mathbf{k}) \omega^2/c_0^2 = (k_\sigma + i/2\ell_\sigma)^2$ that here depend both on circular polarization and wave vector direction.
 In radiative transport, the electric field  in and outside the medium can be imagined to be emitted by a diffuse random secondary source as
 $ E_i(\mathbf{r}) = \int d^3\mathbf{r}_s \,  {G}_{ij}(\mathbf{r},\mathbf{r}_s) s_j(\mathbf{r}_s) $.
 Consequently, the ``ensemble-averaged" field correlation function is
 \begin{eqnarray}\label{GG}
   \langle E_i(\mathbf{r}) \overline{E}_j(\mathbf{r}')\rangle &=& \int d^3 \mathbf{r}_S \int d^3 \mathbf{x}\
   \langle{G}_{ik}(\mathbf{r},\mathbf{r}_s^+) \rangle\langle\overline{{G}}_{lj}(\mathbf{r}',\mathbf{r}_s^-)\rangle  \nonumber \\
  & \times &\left\langle
   s_k\left(\mathbf{r}_s ^+\right) \overline{s}_l\left(\mathbf{r}_s^-\right)\right\rangle
 \end{eqnarray}
with $\mathbf{r}_s^\pm = \mathbf{r}_s \pm \mathbf{x}/2$.   In the far field outside the sphere
$G_{ij}(\mathbf{r}^\pm) \rightarrow (\delta_{ij}-\hat{r}_i\hat{r}_j) \exp(iKr)\exp(ik\hat{\mathbf{r}}\cdot \mathbf{x}/2)/(-4\pi r)$, and this expression simplies
to a manifestation of Huygens' principle,
\begin{equation}\label{ff}
   \langle E_\sigma(\mathbf{r}) \overline{E}_{\sigma'}(\mathbf{r})\rangle  =
   \frac{1}{(4\pi)^2}\int d^3 \mathbf{r}_s  \frac{ e^{-i(K_\sigma - \overline{K}_{\sigma'})b}}{|\mathbf{r}-\mathbf{r}_s|^2}
S_{\sigma \sigma'}(\mathbf{k},\mathbf{r}_s)
\end{equation}
with $b$ the length traversed by the light from the source to its exit point (Figure 2). Because $K_\sigma$ is complex-valued inside the sphere, the integral
is restricted to the skin layer and the
far field correlation function is
proportional to  the Fourier transform $S(\mathbf{k},\mathbf{r}_s)$ of the source correlation function
$\langle s(\mathbf{r}_s^+)\overline{s}(\mathbf{r}_s^-)\rangle$ (in radiative transfer  proportional to the local specific intensity)
with $\mathbf{x}$, integrated over the surface of the sphere.
 A fundamental result from transport theory, is that in the diffuse regime,
 \begin{equation}\label{ss}
   S_{\sigma \sigma'} (\mathbf{k},\mathbf{r}_s) =  2\pi k \delta_{\sigma\sigma'} \mathrm{Im}\, \varepsilon_{\sigma}(\mathbf{k})
   \left(1 - \frac{3 }{c_0}\hat{\mathbf{k}}\cdot \mathbf{D} \cdot \nabla \right) \rho(\mathbf{{r}}_s)
\end{equation}

\begin{figure}
  \includegraphics[width=7cm]{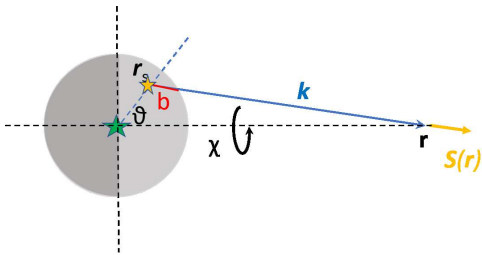}
  \caption{Geometry of the PHE in the far field of the sphere. The diffuse light arriving from the source in the center initiates secondary random sources
  at $\mathbf{r}_{s}$ within a skin layer of depth $\ell$ near the surface, that contribute to the Poynting vector $\mathbf{S}$ in the far field.
  The induced PHE at $\mathbf{r}$ is largest when the magnetic field is perpendicular to the plane.}\label{geometry}
\end{figure}

This is basically a manifestation of the fluctuation dissipation theorem, with the first term standing for detailed balance, and the
gradient term causing the diffuse energy flow.
This source function depends on the magneto-optical birefringence of the effective medium, as well as on
the PHE via
$D_{ij} = D \delta_{ij} + D_H \epsilon_{ijz}$.
In the far field outside the medium the Poynting vector is just $\mathbf{S} =c_0 \hat{\mathbf{k}}(\hat{\mathbf{r}}_s) |E(\mathbf{r})|^2  $, with $\hat{\mathbf{k}}$ the unit vector
along the vector $\mathbf{r}-\mathbf{r}_s$. We thus find
\begin{eqnarray}
  \mathbf{S}\left(\mathbf{r},t+\frac{r}{c_0}\right) &=&\frac{c_0}{8\pi  } \sum_\sigma \int d^2 \mathbf{\hat{r}}_s \int_0^\infty  dz\,  e^{-z/\mu \ell_\sigma} \frac{\hat{\mathbf{k}}(\mathbf{r}_s)}{|\mathbf{r}-\mathbf{r}_s|^2}
  \nonumber \\
     &\times & \frac{1}{\ell_\sigma} \left(1 - \frac{3 }{c_0}\hat{\mathbf{k}}(\mathbf{r}_s)\cdot \mathbf{D} \cdot \nabla \right) \rho(a\hat{\mathbf{r}}_s,t)
\end{eqnarray}
with $\mu = \cos \theta$, $b \approx z/\mu$ and the integral of $\mathbf{r}_S$ extending over half the outer surface of the sphere visible from $\mathbf{r}$. Hence,
\begin{eqnarray}
  \mathbf{S}\left(\mathbf{r},t+\frac{r}{c_0}\right) =\frac{c_0}{4\pi  } \int d^2 \mathbf{\hat{r}}_s  \frac{\mu \hat{\mathbf{k}}(\mathbf{r}_s)}{|\mathbf{r}-\mathbf{r}_s|^2} \nonumber
  \\
     \times  \left(\rho(a,t) - \frac{3 }{c_0}\hat{\mathbf{k}}(\mathbf{r}_s)\cdot \mathbf{D} \cdot \hat{\mathbf{r}}_s \partial_r \rho(a,t)
     \right)
\end{eqnarray}
The only magneto-optical effect that has survived in this expression for the Poynting vector is the PHE contained in $\mathbf{D}$.
The leading radial flow of energy is
obtained by putting $\mathbf{\hat{k}} \approx \mathbf{\hat{r}}$, and becomes
\begin{eqnarray}
  \mathbf{S}_r\left(\mathbf{r},t+\frac{r}{c_0}\right) &=&\frac{c_0a^2}{2 r^2}  \mathbf{\hat{r}} \int_0^1  d\mu \mu
\left(\rho(a,t) - \frac{3 }{c_0} \mu D \partial_r \rho(a,t)\right) \nonumber \\
&=& \frac{c_0a^2}{4 r^2}  \mathbf{\hat{r}}
\left(\rho(a,t) - \frac{2 }{c_0} D \partial_r \rho(a,t)\right)
\end{eqnarray}
The usual radiative boundary condition at the surface is $\rho + \frac{2 }{c_0} D \partial_r \rho =0 $, and identifies $z_0= 2D/c_0$ as the extrapolation
length \cite{skin}. The total current leaving the sphere at distance $r$ is therefore $J(r,t) = 4 \pi r^2 S_r = -4\pi a^2 D \partial_r \rho(a,t-r/c_0)$.
The PHE vanishes in the same approximation $\mathbf{\hat{k}} \approx \mathbf{\hat{r}}$ since for any fixed angle $\theta$, the azimuthal integral  $\int d\chi \, \mathbf{r}_s $ (Fig. 2) is directed along $\mathbf{r}$, and
$\mathbf{\hat{r}}\cdot \mathbf{D}_H \cdot \mathbf{\hat{r}}=0$. Since $\mathbf{\hat{k}}= \mathbf{\hat{r}}(1+ \mu r_s/r) - \mathbf{r}_S/r + \mathcal{O}(1/r^2)$
the PHE generates a Poynting vector in the far field given by
 \begin{eqnarray}
  \mathbf{S}\left(\mathbf{r},t\right) &=& \frac{c_0a^2}{4\pi  r^2}   \int_0^1 d\mu \int_0^{2\pi} d\chi \,   \mu\left( \frac{-\mathbf{r}_S}{r} \right)
 \mathbf{\hat{r}} \cdot ( \mathbf{\hat{r}}_s \times \mathbf{\hat{z}} ) \nonumber \\
 &&  \times \ \frac{-3D_H}{c_0} \partial_r \rho\left(a,t-\frac{r}{c_0}\right) \nonumber \\
 &=& \frac{3}{16}\frac{a^2}{r^3} D_H \partial_r \rho\left(a,t-\frac{r}{c_0}\right) ( \mathbf{\hat{z}} \times \mathbf{\hat{r}} )
\end{eqnarray}
This expression corresponds to an Hall angle outside the sphere equal to $S_\phi /S_r = (3/16)(D_H/D) a/r$ with the same sign as inside the sphere, that decays
slowly with distance. The angular
 momentum carried by the leaking electromagnetic field is
 \begin{eqnarray}\label{Jfinal}
   \mathbf{K} &=&  \frac{1}{c_0^2}\int d^3 \mathbf{r} \ \mathbf{r} \times \mathbf{S}  \nonumber \\
    &=& - \mathbf{\hat{z}}\frac{D_H }{D} \frac{a}{8c_0^2} \int_a^\infty dr \ J\left(a,t-\frac{r}{c_0}\right)
 \end{eqnarray}
 in terms of the total energy current $J(a,t)$ leaving the sphere. It is directed along the vector $-D_H \mathbf{B}_0$ and propagates outwardly.
 This conclusion can also be obtained by considering the righthand side of Eq.~(\ref{AMcon}).
 The total angular momentum grows with time as longs as energy is injected into the sphere.  Since $J(t<0) =0$ the radial integral extends until
 $r= c_0 t$ so that we can identify the torque on the sphere as
  \begin{equation}\label{torquefinal}
  \mathbf{N}= -\frac{d\mathbf{K} }{dt}  = \mathbf{\hat{z}}\frac{D_H }{D} \frac{a}{8c_0} J\left(a,t-\frac{a}{c_0}\right)
  \end{equation}
Equation~(\ref{torquefinal}) is valid for $a \gg \ell$.
It is basically equivalent to the result obtained in Eq.(\ref{torque1}). The extra factor $1/8$  stems from the fact that not all
photons leave the diffuse medium at the
photon Hall angle.
For a photon Hall angle of $10^{-3}$, $a = 1$ mm and $P = 10 $ W, this yields $N= 4\cdot 10^{-15}$ Nm. The observation of such
a value is within experimental reach using a torsion pendulum \cite{pendulum}. Figure 3 shows a possible implementation of
an experiment to observe this effect.

\begin{figure}
  \includegraphics[width=6cm]{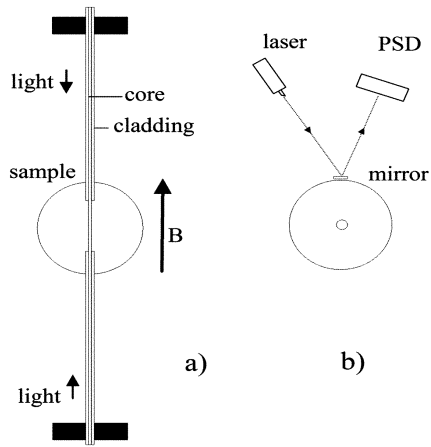}
  \caption{Proposed experimental setup for the observation of the photon Hall pinwheel effect (not to scale). a) side view. The optical fiber serves as torsion
   wire, with the fiber cladding removed near the center of the spherical agglomerate of scatterers (sample) with radius a. The light injected into the two fiber
  ends is chopped at the torsion pendulum resonance frequency. b) Mid-plane top view. Measurement of the torsion angle by deflection of a
  laser onto a position sensitive detector, which is read out with a phase sensitive detector at the light chopping frequency}\label{PHEAMexp}
\end{figure}

Under constant flux during a time $T$, the total angular momentum irreversibly radiated into space
is much bigger than the angular momentum $K \sim  D_H W_S /c_0^2$ derived in Eq.~(\ref{am1}) reversibly taken from the sphere during radiation.
With a moment of inertia  $I =  (8\pi/15)  \rho_m a^5$ the angular velocity of the sphere grows in time as  $\varpi = N \times T/I$. After an exposure time of
one hour, and a mass density $ \rho_M = 1,5  $g/cm$^3$ (corresponding to glycerol doped with a $10\, \%$ volume fraction EuF$_2$ particles at $4,2$ K;
EuF$_2$  has a large Faraday rotation of $1000$ rad/m for a magnetic field of 1 Tesla),
we find  $\varpi = 10$ rad/s.  This is not small at all.
The photon Hall pinwheel effect is  not small either
from the
perspective of a single photon. Equation~(\ref{Jfinal})  tells that the total angular momentum is proportional to the product of $\phi_H (a/c_0) $ and
the total electromagnetic
energy $W$ radiated. If we write $W = n \hbar \omega$ with $n$ the number of injected photons,
the total angular momentum carried away per photon, and measured in units of $\hbar$, is given by
$K/n\hbar = \phi_H ka/8  $. A sphere of $1$ mm  in radius has $ka = 12000$ at optical frequencies. For $\phi_H \approx 10^{-3} $,
each photon leaving the sphere takes away roughly one $ \hbar $ of angular momentum.

\bigskip

The photon Hall pinwheel effect outlined above represents an entire novel class of phenomena, that link diffusive transport in a 
magnetic field to angular momentum. Other diffusive magneto-transport effects, like the ones mentioned in the introduction, should lead to similar pinwheel effects.  
An analogy can be argued to exist between this effect and the Einstein-De Haas effect, that establishes the link between magnetism and angular momentum.
In particular, we conjecture that a star,  in possession of  a strong magnetic field, might be subject to a torque. 
Low energy free electrons are often a major source of radiative transfer  
\cite{Rybicky} via multiple Thomson scattering, which is affected by the magnetic field \cite{japan}.

In summary, we have demonstrated the existence of a circularly rotating Poynting vector, induced by the 
photon Hall effect in a sphere consisting of random magneto-optical scatterers, illuminated by a central optical source, 
and placed in a homogenous magnetic field. This Hall-type photon flux carries a finite orbital angular momentum. 
Surprisingly, the radiation leaving the sphere carries angular momentum with it, and exerts a torque on the sphere. 
Realistic estimates show that the observation of this effect is within experimental reach.

We thank Robert Whitney for drawing attention to the pinwheel effect.

\end{document}